\documentstyle[epsfig,12pt]{article}  
\newcommand{\beq}{\begin{equation}}
\newcommand{\eeq}{\end{equation}}
\newcommand{\bea}{\vspace{0.25cm}\begin{eqnarray}}
\newcommand{\eea}{\end{eqnarray}}

\newcommand{\r}{\mbox{{\boldmath
$\rho$}}}

\newcommand{\qb}{\mbox{{\bf
q}}}
\newcommand{\pb}{\mbox{{\bf
p}}}

\newcommand{\Hb}{\mbox{{\bf
H}}}
\newcommand{\Fb}{\mbox{{\bf
F}}}
\newcommand{\Gb}{\mbox{{\bf
G}}}

\newcommand{\mqb}{\bar{\mbox{{\bf
q}}}}

\newcommand{\fb}{\mbox{{\bf
f}}}
\newcommand{\vb}{\mbox{{\bf
v}}}

\def\lsim{\mathrel{\rlap{\lower4pt\hbox{\hskip1pt$\sim$}}
    \raise1pt\hbox{$<$}}}         %less than or approx. symbol
\def\gsim{\mathrel{\rlap{\lower4pt\hbox{\hskip1pt$\sim$}}
    \raise1pt\hbox{$>$}}}         %greater than or approx. symbol
\begin{document}
\vspace*{-2cm}
 
\bigskip
%%%%%%%%%%%%%%%%%%%%%%%%%%%%%%%%%%%%%%%%%%%%%%%%%%%%%%%%%%
%%%%%%%%%%%%%%%%%%%%%%%%%%%%%%%%%%%%%%%%%%%%%%%%%%%%%%%%%% 

\begin{center}

\renewcommand{\thefootnote}{\fnsymbol{footnote}}

  {\Large\bf
Parton energy loss
due to synchrotron-like gluon emission
\\
\vspace{.7cm}
  }
\medskip
{\large
  B.G.~Zakharov
  \bigskip
  \\
  }
{\it  
 L.D.~Landau Institute for Theoretical Physics,
        GSP-1, 117940,\\ Kosygina Str. 2, 117334 Moscow, Russia
\vspace{1.7cm}\\}

  {\bf
  Abstract}
\end{center}
{
\baselineskip=9pt
We develop a quasiclassical theory of the synchrotron-like gluon radiation. 
Our calculations show that the parton energy loss due to
the synchrotron gluon emission may be important in the jet quenching
phenomenon if the plasma instabilities generate a sufficiently 
strong chromomagnetic field.
Our gluon spectrum disagrees with that
obtained by Shuryak and Zahed within the 
Schwinger's proper time method.
\vspace{.5cm}
\\
}
\noindent{\bf 1}.
The hydrodynamic description \cite{hydro} of the RHIC 
hadron spectra
at low $p_{T}$ gives strong evidence for the thermalization
of the produced hot quark-gluon plasma (QGP)
at the time scale $\tau_{0}\sim 1$ fm.  
The early thermalization provides a serious challenge for the perturbative 
approach based on the Boltzmann equation with the binary and the
$2\leftrightarrow3$ processes \cite{BMSS}. The fast thermalization can probably
be explained if the plasma is strongly coupled \cite{sQGP}.
A promising possibility of speeding the equilibration 
in a weakly coupled plasma scenario is connected with plasma instabilities
\cite{M1,Arnold1,Arnold2,M2,Rebhan1}. One of the potentially important 
instabilities is the 
Weibel \cite{Weibel} instability which occurs for the anisotropic 
plasma distribution. In the QGP produced in $AA$-collisions
the initial parton distribution has $p_{z}\ll
p_{T}$ \cite{BMSS}. In this case the Weibel instability can generate 
considerable (predominantly transverse) chromomagnetic
field which bends the parton trajectories and leads eventually to 
isotropization of the parton distribution \cite{Arnold1,M2,Rebhan1}.
Also, the generated chromomagnetic field may be important
for explanation
of the anomalously small plasma viscosity \cite{BMueller1} and 
the longitudinal broadening of the jet cone (the ridge effect) 
\cite{BMueller2}.

In the scenario with generation of the chromomagnetic field an
interesting question arises on the effect of the chromomagnetic field
on the jet quenching due to the synchrotron-like gluon emission
from fast partons. Evidently the available analyses of the jet
quenching, based on the idea of the induced gluon emission
due to multiple scattering \cite{BDMPS,Z1,GLV1,W1,BSZ},
become inappropriate if the contribution of this mechanism 
to the parton energy loss is comparable to that from
the ordinary multiple scattering. 
It is clear that in this case both 
these mechanism must be treated on an even footing.
As a first step in understanding the role of the chromomagnetic
field in the parton energy loss it would be interesting to calculate the
purely synchrotron induced gluon radiation neglecting the interference between
the two mechanisms. It requires a formalism for the synchrotron radiation 
in QCD.

In the present paper we develop a semiclassical approach to the
synchrotron gluon radiation and give a qualitative estimate of the 
parton energy loss in the QGP produced in $AA$-collisions.
For photon emission our result agrees with prediction of the quasiclassical
operator approach \cite{BK}. 
For gluon emission our spectrum disagrees with that 
obtained by Shuryak and 
Zahed \cite{SZ} within the Schwinger's proper time method.
We give simple physical arguments that the spectrum obtained 
in \cite{SZ} is incorrect.

\vspace{0.2cm}
\noindent{\bf 2}.
We consider the synchrotron gluon radiation from a fast parton
in the quasiclassical regime when for each parton (initial or final) 
the wavelength is much smaller than its Larmor radius, $R_{L}$.
One can show that in this regime, similarly to the photon radiation
in QED, the coherence length of the gluon emission, $L_{c}$, is small 
compared to the minimal $R_{L}$.
It allows one to perform the calculation of the radiation rate per unit 
length by considering the case of a slab of chromomagnetic field of 
thickness $L$ which is large compared to $L_{c}$, but small compared to 
the minimal $R_{L}$. In this case the transverse momenta of the final partons
are small compared to their longitudinal momenta (we choose the $z$-axis 
along the initial parton momentum). We consider the case of a slab
perpendicular to the $z$-axis with transverse chromomagnetic field, $\Hb_{a}$.
It is enough to consider the chromomagnetic field with the only nonzero 
color components in the Cartan subalgebra, i.e., for $a=3$ and $a=8$ for 
the $SU(3)$ color group. The interaction of the gluons with the background
chromomagnetic field is diagonalized by introducing the fields having definite
color isospin, $Q_{A}$, and color hypercharge, $Q_{B}$, (we will describe 
the color charge by the two-dimensional vector $Q=(Q_A,Q_B)$).
In terms of the usual gluon vector potential, $G$, the diagonal color gluon
states read
(the Lorentz indices are omitted)
$X=(G_{1}+iG_{2})/\sqrt{2}$ ($Q=(-1,0)$),
$Y=(G_{4}+iG_{5})/\sqrt{2}$ ($Q=(-1/2,-\sqrt{3}/2)$),
$Z=(G_{6}+iG_{7})/\sqrt{2}$ ($Q=(1/2,-\sqrt{3}/2)$).
The neutral gluons $A=G_{3}$ and $B=G_{3}$ with $Q=(0,0)$,
to leading order in the coupling
constant, do not interact with the background field, and the emission of
these gluons are similar to the photon radiation in QED.

The $S$-matrix element of the $q\rightarrow gq'$ synchrotron transition
can be written as (we omit the color factors and indices)
\beq
\langle gq'|\hat{S}|q\rangle=-ig\int\! dy 
\bar{\psi}_{q'}(y)\gamma^{\mu}G_{\mu}^{*}(y)\psi_{q}(y)\,,
\label{eq:10}
\eeq
where $\psi_{q,q'}$ are the wave 
functions of the initial quark and final quark, $G$ is the wave function
of the emitted gluon.
We write each quark wave function in the form
$
\psi_{i}(y)=\exp[-iE_{i}(t-z)]
\hat{u}_{\lambda}
\phi_{i}(z,\r)/\sqrt{2E_{i}}$ (hereafter the bold vectors denote the
transverse
vectors), where $\lambda$ is quark helicity,
$\hat{u}_{\lambda}$ is the Dirac spinor operator.
The $z$-dependence of the transverse wave functions
$\phi_{i}$ is governed by the two-dimensional 
Schr\"odinger equation 
\beq
i\frac{\partial\phi_{i}(z,\r)}{\partial
z}={\Big\{}\frac{
(\pb-gQ_{n}\Gb_{n})^{2}
+m^{2}_{q}} {2E_{i}}
+gQ_{n}(G^{0}_{n}-G^{3}_{n}){\Big\}}
\phi_{i}(z,\r)\,,
\label{eq:20}
\eeq
where now $G$ denotes
the external vector potential (the superscripts are the Lorentz indexes and 
$n=1,2$ correspond to the $A$ and $B$ 
color components in the Cartan subalgebra), $Q_{n}$ is the quark color charge. 
The wave function of the emitted
gluon can be represented in a similar way.
We will assume that in the QGP for the usual parton masses 
one can use the corresponding quasiparticle masses.

We take the external vector potential in the form 
$G^{3}_{n}=[\Hb_{n}\times \r]^{3}$,
$\Gb_{n}=0$, $G^{0}_{n}=0$ (we assume that chromoelectric field is absent, 
however,
it can be included as well). For this choice of the vector potential
the term $-gQ_{n}G^{3}_{n}$ in (\ref{eq:20}) can be viewed as the
potential energy in the impact parameter plane $U_{i}=-\Fb_{i}\cdot\r$,
where $\Fb_{i}$ is the corresponding Lorentz force. Then 
the solution of (\ref{eq:20}) can be taken in the form
\beq
\phi_{i}(z,\r)=\exp{\left\{i\pb_{i}(z)\r-
\frac{i}{2E_{i}}\int_{0}^{z}dz'[\pb^{2}_{i}(z')+m^{2}_{q}]\right\}}\,.
\label{eq:30}
\eeq
Here the transverse momentum $\pb_{i}(z)$ is the solution to 
the parton equation of motion
in the impact parameter plane
\beq
\frac{d\pb_{i}}{dz}=\Fb_{i}(z)\,.
\label{eq:40}
\eeq
Below  we denote the value of 
$\pb_{i}(\pm\infty)$ 
as $\pb_{i}^{\pm}$. 
By using (\ref{eq:10}), (\ref{eq:30}) one can obtain 
\bea
\langle gq'|\hat{S}|q\rangle=-ig
(2\pi)^{3}\delta(E_{g}+E_{q'}-E_{q})
\int_{-\infty}^{\infty}dz V(z,\{\lambda\})
\delta(\pb_{g}(z)+\pb_{q'}(z)-\pb_{q}(z))\nonumber\\
\times
\exp{\left\{-i\int_{0}^{z} dz'
\left[\frac{\pb^{2}_{q}(z')+m^{2}_{q}}{2E_{q}}
-\frac{\pb^{2}_{g}(z')+m^{2}_{g}}{2E_{g}}
-\frac{\pb^{2}_{q'}(z')+m^{2}_{q}}{2E_{q'}}
\right]\right\}}\,,
\label{eq:50}
\eea
where $V$ is the spin vertex factor, $\{\lambda\}$ is the set of the parton
helicities. For the transition conserving
quark helicity 
$V=-iE_{q}\sqrt{1-x}[2\lambda_{q}x+(2-x)\lambda_{g}]
[v_{x}(z)-i\lambda_{g}v_{y}(z)]/\sqrt{2}$, and for the spin-flip
case $V=ixm_{q}(2\lambda_{q}\lambda_{g}+1)/\sqrt{2(1-x)}$. Here
$\vb(z)=\vb_{g}(z)-\vb_{q}(z)$ is the
relative transverse velocity in the final $gq'$ parton system
which can be written as 
$\vb(z)=\qb(z)/\mu$ with $\qb(z)=\pb_{g}(z)(1-x)-\pb_{q'}(z)x$,
$\mu=E_{q}x(1-x)$ (here $x$ is the longitudinal gluon fractional momentum).
Due to 
the color charge conservation $\Fb_q=\Fb_g+\Fb_{q'}$. For this reason,
the argument of the second $\delta$-function does not depend on $z$,
and can be replaced by $\pb_{g}^{+}+\pb_{q'}^{+}-\pb_{q}^{+}$. 
From (\ref{eq:50}) we obtain the gluon emission spectrum
\bea
\frac{dP}{dx}=
\frac{1}{(2\pi)^{2}}
\int d\pb_{g}^{+}
\int
dz_{1}dz_{2}
g(z_{1},z_{2})\nonumber\\
\times
\exp{\left\{i\int_{z_{1}}^{z_{2}} dz
\left[\frac{\pb^{2}_{q}(z)+m^{2}_{q}}{2E_{q}}
-\frac{\pb^{2}_{g}(z)+m^{2}_{g}}{2E_{g}}
-\frac{\pb^{2}_{q'}(z)+m^{2}_{q}}{2E_{q'}}
\right]\right\}}\,,
\label{eq:60}
\eea
where the vertex factor reads (we recover the vertex color factor $C$)
\beq
g(z_{1},z_{2})=\frac{C\alpha_{s}}{8E_{q}^{2}x(1-x)}\sum_{\{\lambda\}}
V^{*}(z_2,\{\lambda\})V(z_1,\{\lambda\})=
g_{1}\vb(z_2)\vb(z_1)+g_{2}
\label{eq:70}
\eeq
with $g_{1}=C\alpha_{s}(1-x+x^{2}/2)/x$ and
$g_{2}=C\alpha_{s}m_{q}^{2}x^{3}/2\mu^{2}$ 
(the two terms in (\ref{eq:70}) correspond to the
non-flip and spin-flip processes). 
The color factor reads
$C=|\lambda_{fi}^{a}\chi_{a}^{*}/2|^{2}$, where $i,f$ are the color 
indexes of the initial and final quarks, $\chi_{a}$ is the color wave 
function of the emitted gluon.

For a uniform external field we can write
$\vb(z_2)\vb(z_1)=[\mqb^{2}-\fb^{2}\tau^{2}/4]/\mu^{2}$, where
$\mqb=\qb(\bar{z})$, $\bar{z}=(z_{1}+z_{2})/2)$, $\tau=z_{2}-z_{1}$,
and $\fb=d\qb/dz=\Fb_{g}(1-x)-\Fb_{q'}x$.
The argument of the exponential function in (\ref{eq:60}) 
can be rewritten as
\beq
\Phi(\tau,\mqb)=\frac{(\epsilon^{2}+\mqb^{2})\tau}{2\mu}
+\frac{\fb^{2}\tau^{3}}{24\mu}\,
\label{eq:80}
\eeq
with $\epsilon^{2}=m_{q}^{2}x^{2}+m_{g}^{2}(1-x)$.
After replacing in (\ref{eq:60}) the 
integration over $\pb_{g}^{+}$ by the integration over $\bar{\qb}$
we obtain for the radiation rate per unit length
\bea
\frac{dP}{dLdx}=
\frac{1}{(2\pi)^{2}}
\int d\mqb
\int_{-\infty}^{\infty} d\tau
\left[\frac{g_{1}}{\mu^{2}}
\left(\mqb^{2}-\frac{\fb^{2}\tau^{2}}{4}\right)
+g_{2}\right]
\exp{[-i\Phi(\tau,\mqb)]}\,.
\label{eq:90}
\eea
With the help of integration by parts one can rewrite (\ref{eq:90}) as
\bea
\frac{dP}{dLdx}=
-\frac{1}{(2\pi)^{2}}
\int d\mqb
\int_{-\infty}^{\infty} d\tau
\left[\frac{g_{1}}{\mu^{2}}\left(\epsilon^{2}+\frac{\fb^{2}\tau^{2}}{2}\right)
-g_{2}\right]
\exp{[-i\Phi(\tau,\mqb)]}\,.
\label{eq:100}
\eea
After integrating over $\mqb$ (\ref{eq:100}) takes the form 
\beq
\frac{dP}{dLdx}=
\frac{i\mu}{2\pi}
\int_{-\infty}^{\infty} \frac{d\tau}{\tau}
\left[\frac{g_{1}}{\mu^{2}}\left(\epsilon^{2}+\frac{\fb^{2}\tau^{2}}{2}\right)
-g_{2}\right]
\exp{\left\{-i\left[
\frac{\epsilon^{2}\tau}{2\mu}
+\frac{\fb^{2}\tau^{3}}{24\mu}\right]\right\}}\,.
\label{eq:110}
\eeq
Note that in (\ref{eq:90})-(\ref{eq:110}) it is assumed that $\tau$ has a 
small negative imaginary part.  One can easily show that in (\ref{eq:110}) 
the integral around the lower semicircle near the pole at $\tau=0$ plays 
the role of the $\fb=0$ subtraction term. 
Expressing the integrals along the real axis in terms of the Airy function
$\mbox{Ai}(z)=\frac{1}{\pi}\sqrt{\frac{z}{3}}K_{1/3}(2z^{3/2}/3)$ (here
$K_{1/3}$ is the Bessel function) (\ref{eq:110}) can be written as
\beq
\frac{dP}{dLdx}=
\frac{a}{\kappa}\mbox{Ai}^{'}(\kappa)+
b\int_{\kappa}^{\infty}dy\mbox{Ai}(y)\,,
\label{eq:120}
\eeq
where 
$a=-{2\epsilon^{2}g_{1}}/{\mu}$,
$
b=\mu  g_{2}-{\epsilon^{2}g_{1}}/{\mu}
$,
$\kappa=\epsilon^{2}/(\mu^{2}\fb^{2})^{1/3}$.

From (\ref{eq:110}) one can obtain for the coherence length of 
the gluon emission
$L_{c}\sim \mbox{min}(L_{1},L_{2})$, where $L_{1}=2\mu/\epsilon^{2}$ and
$L_{2}=(24\mu/\fb^{2})^{1/3}$. From this estimate one can easily show that for 
radiation of a charged gluon the condition $L_{c}/R_{g,L}\ll 1$ (which
is necessary for validity of our small angle approximation) is really
satisfied in the quasiclassical regime when $E_{g}R_{g,L}\gg 1$ and 
$E_{g}\gg m_{g}$. In the interesting to us region $F_{q'}\lsim m_{g}^2$
(hereafter $F_{i}=|\Fb_{i}|$) 
for emission of a neutral gluon the condition
$L_{c}/R_{q,L}\ll 1$ is also satisfied.

Our spectrum for neutral gluons when $F_{g}=0$ for $m_{g}=0$ agrees with 
the photon spectrum obtained in the quasiclassical operator approach
\cite{BK}\footnote{Note that for the photon emission by an electron
the integrand in the exponential term in (\ref{eq:60}) can also 
be written as $E_{e}k^{\mu}x_{\mu}(t)/(E_{e}-E_{\gamma})$, where $k^{\mu}$ 
is the photon four momentum and $x_{\mu}(t)$ is the classical electron
trajectory. This gives precisely the spectrum in the form obtained
by Baier and Katkov \cite{BK}. 
}, 
and  similarly to the photon emission (\ref{eq:120}) gives 
$dP/dLdx\propto x^{-2/3}$ at $x\ll 1$. The nonzero gluon mass leads to 
the Ter-Mikaelian suppression
at $x\lsim (m_{g}^{3}/E_{q}F_{q'})^{1/2}$. In this region the parameter $\kappa$
becomes larger than unity and the spectrum
$dP/dLdx\propto \exp{[-2m_{g}^{3}/3E_{q}F_{q'}x^{2}]}/x$.
For the charged gluons from (\ref{eq:120}) one can obtain in the massless
limit  $dP/dLdx\propto x^{-4/3}$ at $x\ll 1$. The Ter-Mikaelian mass effect
suppresses the gluon spectrum at $x\lsim m_{g}^{3}/E_{q}F_{g}$.
In this region $dP/dLdx\propto \exp{[-2m_{g}^{3}/3xE_{q}F_{g}]}/x^{3/2}$
at $x\ll 1$. Note that in general if one neglects the parton masses
(or in the strong field limit) the spectrum (\ref{eq:120}) takes a simple
form
\beq
\frac{dP}{dLdx}\approx
\frac{\alpha_{s}C\Gamma(2/3)[1-x+x^{2}/2](9\fb^{2})^{1/3}}
{\pi\sqrt{3} x[E_{q}x(1-x)]^{1/3}}\,.
\label{eq:121}
\eeq

Our derivation is valid
for the $g\rightarrow gg$ transition as well. In this case 
$a$ and $b$ have the same form but with $\epsilon^{2}=m_{g}^{2}(1-x+x^{2})$, 
$g_{1}=C\alpha_{s}[1+x^{4}+(1-x)^{4}]/4x(1-x)$ and $g_{2}=0$.
The gluon vertex color factor reads 
$C=|\chi_{a}\chi_{b}^{*}\chi_{c}^{*}f_{abc}|^{2}$, where the index 
$a$ corresponds to the initial gluon, and 
$b$, $c$ to the final gluons.

\vspace{0.2cm}
\noindent{\bf 3}.
In our formula for the spectrum for $a\rightarrow bc$
transition  all the properties of the final $bc$ state are 
only accumulated
in 
$\fb^{2}=\Fb_{b}^{2}x_{b}^{2}-2\Fb_{b}\Fb_{c}x_{b}x_{c}+\Fb_{c}^{2}x_{c}^{2}$ 
(except for the trivial vertex factor). The $\fb^{2}$ crucially
depends on the relation between the forces acting on the final partons.
The value of $\fb^{2}$ characterizes the difference in bending of the
trajectories of the $b$ and $c$ partons in the external field which is
responsible for the synchrotron radiation.
For this reason the spectrum vanishes if at some $x$ $\fb=0$. 
This, for example,
occurs for the $g_{X}\rightarrow g_{Y}g_{\bar{Z}}$ gluon process 
at $x=0.5$ for the 
external field in the color state $A$.

Note, that the spectrum obtained in the present paper can also
be derived making use the light-cone path integral formalism \cite{Z1}
for gluon emission due to multiple scattering. In \cite{Z1} the spectrum
was expressed in terms of the Green's function for the Schr\"odinger 
equation describing the relative motion in the $\bar{q}q'g$ system.
In the absence of the external field 
the corresponding Hamiltonian has a purely imaginary potential 
$U(\r)=-i n\sigma_{\bar{q}q'g}(|\r|)/2$ (here
$\sigma_{\bar{q}q'g}$ is the cross section 
for the $\bar{q}q'g$ state, and $n$ is the number density of the medium).  
In the case of the synchrotron radiation (without multiple scattering)
this potential should be replaced by the real potential $U(\r)=-\fb\cdot \r$.
The advantage of the path integral approach is that it allows one to 
treat the gluon emission including both the multiple scattering
and bending of the trajectories in the external field.
This analysis will be given in further publications.

Our formula (\ref{eq:110}) disagrees with that obtained 
by Shuryak and Zahed in the soft gluon limit within the Schwinger's
proper time method.
\cite{SZ}. In the 
spectrum derived in \cite{SZ} 
(Eq. (20) of \cite{SZ}) the argument of the exponential  contains 
(we use our notation) 
$\Fb_{q'}^{2}x_{g}^{2}+\Fb_{g}^{2}$
instead of our 
$\fb^{2}$. Also, in the
pre-exponential factor instead of $\fb^{2}$ there appears 
$\Fb_{q'}^{2}x_{g}^{2}$.  
Thus, our result at $x_{g}\ll 1$ agrees with that of \cite{SZ} 
only for the QED like
processes (emission of the neutral gluons). 
For a real QCD process, with charged gluon and quark, due to 
the absence of the interference term the spectrum of \cite{SZ} 
is insensitive to the relation between the signs of the color charges 
of the final partons. It is strange enough,
since the difference
in the bending of the final parton trajectories
(which is responsible for the synchrotron radiation) is sensitive
to the relation between the color charges of the final partons.
Also, Eq. (20) of \cite{SZ} gives clearly wrong prediction that in 
the massless limit
the spectrum of the $q_{1}\rightarrow g_{Z}q_{3}$ transition
for the chromomagnetic field in the color state $A$  
vanishes (since in this case $\Fb_{q'}=0$). Indeed,
this process except for the spin effects is analogous to the 
synchrotron radiation in QED, and there is no physical reason 
why it should vanish.
Note also that since in \cite{SZ} the pre-exponential factor contains
the Lorentz forces acting on the final partons 
in non-symmetric form it is clear that for the $g\rightarrow gg$
process the method of \cite{SZ} should give  the spectrum with
incorrect permutation properties
Thus, one sees that the formula obtained in \cite{SZ} clearly
leads to physically absurd predictions. Unfortunately, the details 
of the calculations have not been given in \cite{SZ}. For
this reason it is difficult to understand what is really wrong
in the analysis \cite{SZ}. 

\vspace{0.2cm}
\noindent{\bf 4.}
We use the quasiparticle masses
obtained from the 
analysis of the lattice data within the quasiparticle model \cite{LH}. 
For the plasma temperature $T\sim (1-3)T_{c}$ which is relevant to
the RHIC and LHC conditions the analysis 
\cite{LH} gives $m_{q}\approx 0.3$ and $m_{g}\approx 0.4$ GeV. 
In Figs.~1,~2 we present the averaged over the color states gluon spectra 
for $q\rightarrow  g q'$ and $g\rightarrow  g g$ processes for the 
chromomagnetic field in the color state $A$ for different initial parton 
energies. The computations are performed for $\alpha_{s}=0.3$ and
$gH_{A}/m_{D}^{2}=0.05$, 0.25 and 1, 
where $m_{D}$ is the Debye mass
(we assume that as for an isotropic
weakly coupled plasma $m_{D}^{2}=2m_{g}^{2}$).
The results for the chromomagnetic field
in the color state $B$ are very close to that shown in Figs.~1,~2, and
we do not show them.
The decrease of the spectra at $x\rightarrow 0$ (and $x\rightarrow 1$ for
$g\rightarrow gg$ process) which is well seen for the smallest value of the
field is due to the Ter-Mikaelian mass effect. This suppression decreases 
with increase of the chromomagnetic field.

In Fig.~3 we show the spectra 
for the $q\rightarrow  g q'$ process for different
quark and gluon color states. To illustrate better namely the dependence
on the color indexes in all the spectra we replace
the color vertex factor $C$ by unity. One can see that the spectra grow
strongly in the region of small $x$ with the gluon color charge.
For this reason the averaged over color spectra shown in Figs.~1,~2 
are dominated by the processes with charged gluons.

For understanding the potential role of the synchrotron-like radiation in the 
jet quenching in $AA$-collisions it is interesting to
estimate the energy loss due to the synchrotron gluon emission.
Of course, a realistic estimate of this effect requires detailed information
on the time evolution of the QGP instabilities. Also, as we said in the
introduction, an accurate analysis should treat the synchrotron radiation 
and usual bremsstrahlung due to multiple
scattering on an even footing. This is, however, beyond the scope of the
present analysis. In the present paper we can give only a crude 
estimate of the effect. To estimate the chromomagnetic field 
we rely on the idea that the small viscosity of the QGP observed
at RHIC is due to parton rescatterings in the turbulent magnetic
field \cite{BMueller1}. In this model the viscosity/entropy ratio
$\eta/s\sim 1/g^{2}\xi^{3/2}$ \cite{BMueller1}, where
$\xi$ is the anisotropy parameter of the initial plasma distribution. 
It is expected that the generated magnetic field is saturated at 
$g^{2}\langle H^{2} \rangle \sim \xi^{2} m_{D}^{2}$. To obtain a realistic
$\eta/s$ ratio one should assume that $\xi\sim 1$. Probably, the value of the
magnetic field obtained in this way may be viewed as an upper bound.
Indeed, in this case 
the ratio of the magnetic energy to the thermal parton energy is $\sim 0.3$,
and a higher fraction of the magnetic energy looks unrealistic.
Making use the estimated magnetic field for RHIC 
conditions for $\alpha_{s}=0.3$ we obtained $\Delta E/E\sim 0.1-0.2$
for quarks and $\Delta E/E\sim 0.2-0.4$ for gluons at $E\sim 10-20$ GeV
(for $\alpha_{s}=0.5$ the results
are about two times bigger). 
These estimates have been obtained assuming that the parton path length
in the magnetic field is $\sim 2-4$ fm. Also, we  neglected any finite-size 
effects. These effects may be important if $L_{c}$ and $L$ are 
of the same order. For RHIC conditions the dominating contribution
to the energy loss comes from the soft gluon emission 
where $L_{c}\sim 1-2$ fm. In this situation the finite-size effects
may suppress the energy loss by a factor $\sim 0.5$.
One more suppression mechanism
may be connected with the finite coherence length of the
turbulent magnetic field, $L_{m}$. However, if
for the unstable magnetic field modes the wave vector 
$k^{2}\lsim \xi m_{D}^{2}$ \cite{BMueller1}, 
for RHIC conditions this suppression should not be very strong since
we have $L_{m}/L_c\gsim 1$. In this regime the turbulent effects
should not suppress strongly the energy loss, and  as a plausible estimate
one can take the turbulent suppression factor $\sim 0.5$. Even with
these suppression factors the synchrotron energy loss 
turns out to be comparable with that due to ordinary bremsstrahlung,
and its relative contribution may be larger than that from
the collisional energy loss \cite{Z2007}. 
Of course, the above estimates are very crude. Nevertheless
they demonstrate that the synchrotron radiation can be important in jet 
quenching and deserves further more accurate investigations.
Of course, the synchrotron radiation and  bremsstrahlung are 
not additive, since magnetic bending of the parton trajectories 
will suppress bremsstrahlung due to multiple scattering and vice versa 
multiple scattering will suppress 
the synchrotron radiation.
This, in principle, can mask the effect of the synchrotron emission
in the energy loss.
However, the synchrotron radiation can reveal itself in the longitudinal
broadening of the jet cone. Note that for the synchrotron emission
this effect appears already at the level of the gluon emission itself
contrary to the mechanism of Ref. \cite{BMueller2} where it is
a purely final-state interaction effect.
   
\vspace{.2cm}
\noindent {\bf 5}. 
In summary, we have developed a quasiclassical theory of the 
synchrotron-like gluon radiation. 
In the QGP the gluon spectrum is dominated by the processes with
emission of the charged gluons, the effect of the neutral gluons
is relatively small.
Our calculations show that the parton energy loss due to
the synchrotron radiation may be important in the jet quenching
if the QGP instabilities generate magnetic field
$\langle H^{2} \rangle \sim m_{D}^{2}/g^{2}$.
Our gluon spectrum disagrees with that
obtained by Shuryak and Zahed \cite{SZ}.
We give simple physical arguments that the spectrum derived in \cite{SZ} 
is incorrect.

\vspace {.7 cm}
\noindent
{\large\bf Acknowledgements}

\noindent
This research is supported 
in part by the grant RFBR
06-02-16078-a and the program SS-3472.2008.2.

\newpage
%
%------------------------------------------------------------------
\begin{center}
{\Large \bf Figures}
\end{center}
%------------------------------------------------------------------
%
%\vspace{-2cm}
\begin{figure} [h]
\begin{center}
\hspace{-1cm}\epsfig{file=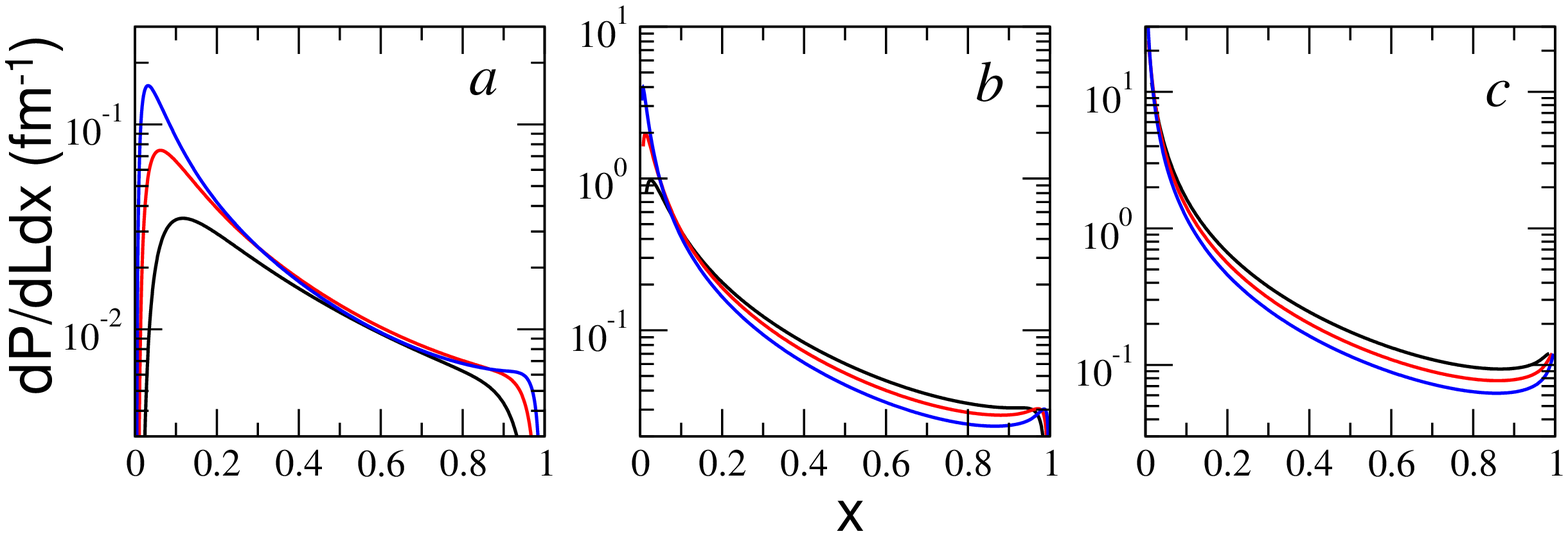,height=14cm,angle=0}
\end{center}
\vspace{-6cm}
\caption[.]
{
The spectrum for the $q\rightarrow gq$ process in the 
chromomagnetic field in the color state $A$ for $\alpha_{s}=0.3$,
$gH_{A}/m_{D}^{2}=0.05$ (a), 0.25 (b) and 1 (c), for the initial quark
energies $E_{q}=20$ GeV (black),
$E_{q}=40$ GeV (red),  
$E_{q}=80$ GeV (blue).
}
\end{figure}

\begin{figure}[t]
\begin{center}
\epsfig{file=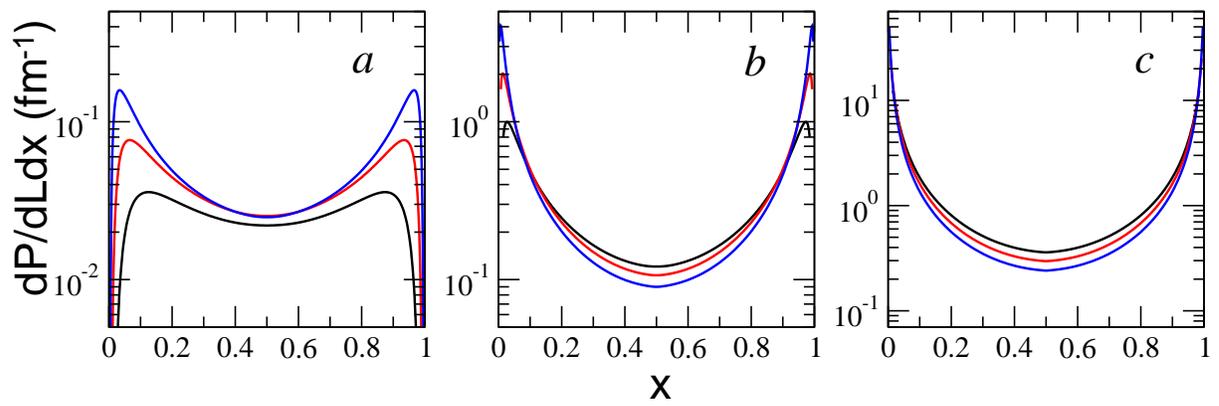,height=14cm,angle=0}
\end{center}
\vspace{-5cm}
\caption[.]
{
The same as in Fig.~1 but for the $g\rightarrow gg$ process.
}
\end{figure}

\begin{figure}[t]
\begin{center}
\epsfig{file=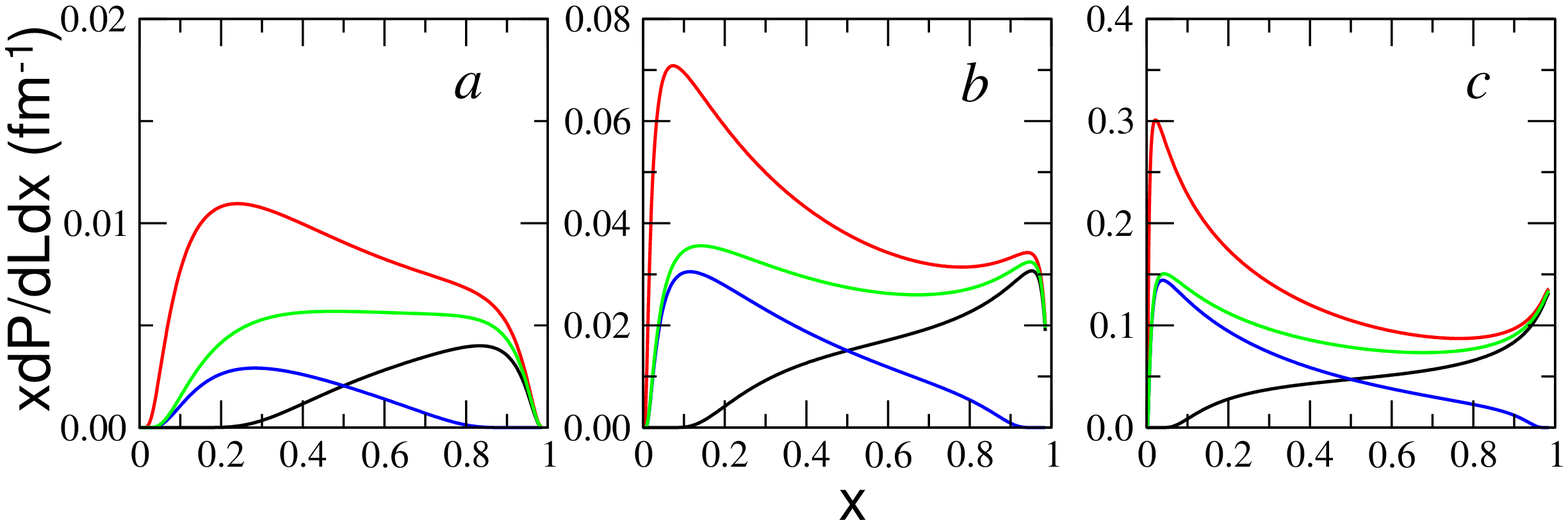,height=14cm,angle=0}
\end{center}
\vspace{-5cm}
\caption[.]{
The spectrum for the $q\rightarrow gq$ process for specific color
states. 
The curves correspond to $q_{1}\rightarrow g_{A}q_{1}$ (black), 
$q_{1}\rightarrow g_{\bar{X}}q_{2}$ (red),
$q_{1}\rightarrow g_{Z}q_{3}$ (blue), 
$q_{3}\rightarrow g_{Y}q_{1}$ (green). 
In all the processes
the vertex color factor $C$ in (\ref{eq:70}) is replaced by unity. 
The computations are performed for $\alpha_{s}=0.3$,
and $E_{q}=20$ GeV. 
The chromomagnetic field is in the color state $A$ with
$gH_{A}/m_{D}^{2}=0.05$ (a), 0.25 (b) and 1 (c).
}
\end{figure}


\begin{thebibliography}{99}

\bibitem{hydro}
% Hydrodynamic description of ultrarelativistic heavy ion collisions.
%Peter F. Kolb (SUNY, Stony Brook) , 
%Ulrich W. Heinz (Ohio State U.) . SUNY-NTG-03-06, May 2003. 82pp. 
%Invited review for 'Quark Gluon Plasma 3'. 
%Editors: R.C. Hwa and X.N. Wang, World Scientific, Singapore.
%In *Hwa, R.C. (ed.) et al.: Quark gluon plasma* 634-714.
%e-Print: nucl-th/0305084
P.F.~Kolb and U.W.~Heinz,
nucl-th/0305084.


\bibitem{BMSS}
% 'Bottom up' thermalization in heavy ion collisions.
%R. Baier (Bielefeld U.) , Alfred H. Mueller (Columbia U.) , 
%D. Schiff (Orsay) , D.T. Son (Columbia U. & RIKEN BNL) . Sep 2000. 10pp.
R.~Baier, A.H.~Mueller, D.~Schiff, and D.T.~Son,
Phys. Lett. B{\bf 502}, 51 (2001) [hep-ph/0009237].

\bibitem{sQGP}
%What RHIC experiments and theory tell us about properties 
%of quark-gluon plasma?
E.V.~Shuryak,
Nucl. Phys. A{\bf 750}, 64 (2005) [hep-ph/0405066].


\bibitem{M1}
S.~Mr\'owczy\'nski,
Phys. Lett. B{\bf 214} 587 (1988); {\it ibid.}
%Color filamentation in ultrarelativistic heavy ion collisions.
%Stanislaw Mrowczynski (Soltan Inst., Swierk) . Jun 1996. 4pp.
B{\bf 393}, 26 (1997);
%e-Print: hep-ph/9606442
%
% Color collective effects at the early stage of 
%ultrarelativistic heavy ion collisions.
%S. Mrowczynski (Warsaw, Inst. Nucl. Studies) . 1994.
%Published in Phys.Rev.C49:2191-2197,1994.
Phys. Rev. C{\bf 49}, 2191 (1994).


\bibitem{Arnold1}
% Quark-gluon plasma thermalization and plasma instabilities.
%Peter Arnold (Virginia U.) . Aug 2004. 10pp.
%Talk given at 6th Conference on Strong and Electroweak 
%Matter 2004 (SEWM04), Helsinki, Finland, 16-19 Jun 2004.
%e-Print: hep-ph/0409002
P.~Arnold, hep-ph/0409002.

\bibitem{Arnold2}
% QCD plasma instabilities: The NonAbelian cascade.
%Peter Arnold (Virginia U.) , Guy D. Moore (McGill U.) . Sep 2005. 20pp.
%Published in Phys.Rev.D73:025006,2006.
%e-Print: hep-ph/0509206
P.~Arnold and G.D.~Moore,
Phys. Rev. D{\bf 73}, 025006 (2006) [hep-ph/0509206].


\bibitem{M2}
% Early stage thermaliza tion via instabilities.
%Stanislaw Mrowczynski (Swietokrzyska Acad & Warsaw, 
%Inst. Nucl. Studies) . Nov 2006. 23pp.
%Presented at 3rd International Workshop on Critical 
%Point and Onset of Deconfinement, Florence, Italy, 3-6 Jul 2006.
%Published in PoS CPOD2006:042,2006.
%e-Print: hep-ph/0611067
S.~Mr\'owczy\'nski, hep-ph/0611067.




\bibitem{Rebhan1}
% Instabilities of an anisotropically expanding 
%non-Abelian plasma: 1D+3V discretized hard-loop simulations.
%Anton Rebhan (Vienna, Tech. U.) , 
%Michael Strickland (Frankfurt U., FIAS & Santa Barbara, KITP) , 
%Maximilian Attems (Vienna, Tech. U.) . 
%TUW-08-05, NSF-KITP-08-01, Feb 2008. 27pp.
%e-Print: arXiv:0802.1714 [hep-ph]
A.~Rebhan,
M.~Strickland, and M.~Attems, hep-ph/0802.1714.



\bibitem{Weibel}
E.S.~Weibel,
Phys. Rev. Lett. {\bf 2}, 83 (1959).



\bibitem{BMueller1}
% Anomalous transport processes in anisotropically expanding 
%quark-gluon plasmas.
%Masayuki Asakawa (Osaka U.) , Steffen A. Bass (Duke U.) , 
%Berndt Muller (Duke U. & Kyoto U., Yukawa Inst., Kyoto) . Aug 2006. 31pp.
%Published in Prog.Theor.Phys.116:725-755,2007.
%e-Print: hep-ph/0608270
M.~Asakawa, S.A.~Bass, and B.~Muller,
Prog. Theor. Phys. {\bf 116}, 725 (2007) [hep-ph/0608270].

\bibitem{BMueller2}
% Longitudinal Broadening of Quenched Jets in Turbulent Color Fields.
A.~Majumder, B.~Muller, and S.A.~Bass,
Phys. Rev. Lett. {\bf 99}, 042301 (2007) [hep-ph/0611135].

\bibitem{BDMPS}
R.~Baier, Y.L.~Dokshitzer, A.H.~Mueller, S.~Peign\'e, and D.~Schiff,
Nucl.\ Phys.\ B{\bf 483}, 291 (1997); {\it ibid.} B{\bf 484}, 265 (1997);
%
R.~Baier, Y.L.~Dokshitzer, A.H.~Mueller, and D.~Schiff,
Nucl.\ Phys.\ B{\bf 531}, 403 (1998).

\bibitem{Z1}
B.G.~Zakharov, JETP\ Lett. {\bf 63}, 952 (1996); {\it ibid.}
{\bf 65}, 615 (1997); {\it ibid.}
{\bf 70}, 176 (1999); 
Phys.\ Atom.\ Nucl. {\bf 61}, 838 (1998).



%REACTION OPERATOR APPROACH TO NONABELIAN ENERGY LOSS.
\bibitem{GLV1}
M.~Gyulassy, P.~L\'evai and I.~Vitev, 
Nucl.\ Phys. B{\bf 594}, 371 (2001).
%(hep-ph/0006010);

%JET QUENCHING VERSUS JET ENHANCEMENT: A QUANTITATIVE STUDY OF 
%THE BDMPS-Z GLUON RADIATION SPECTRUM.
\bibitem{W1}
U.A.~Wiedemann,
Nucl.\ Phys.\ A{\bf 690}, 731 (2001).
%[HEP-PH 0008241]

\bibitem{BSZ}
R.~Baier, D.~Schiff, and B.G.~Zakharov, 
Ann.\ Rev.\ Nucl.\ Part. {\bf 50}, 37 (2000) [hep-ph/0002198].

\bibitem{BK}
V.N.~Baier and V.M.~Katkov, JETP {\bf 26}, 854 (1968).

\bibitem{SZ}
% Jet quenching in high-energy heavy ion collisions 
%by QCD synchrotron - like radiation.
E.V.~Shuryak and I.~Zahed,
Phys. Rev. D{\bf 67}, 054025 (2003) [hep-ph/0207163].

\bibitem{LH}
P.~L\'evai and U.~Heinz,
Phys.\ Rev.\ C{\bf 57}, 1879 (1998).

% Parton energy loss in an expanding quark-gluon plasma: 
%Radiative versus collisional.
\bibitem{Z2007}
B.G.~Zakharov,
JETP Lett. {\bf 86}, 444 (2007) [hep-ph/0708.0816].

\end{thebibliography}
\end{document}